\documentclass[epj]{svjour}
\usepackage{graphicx,bm}
\begin{document}

\title{Shell-model half-lives for r-process $N=82$ nuclei}

\author{J.~J. Cuenca-Garc\'ia\inst{1} \and
  G. Mart\'{\i}nez-Pinedo\inst{1} \and K. Langanke\inst{1,2} \and
  F. Nowacki\inst{3} \and I. N. Borzov\inst{1}}

\institute{Gesellschaft f\"ur Schwerionenforschung Darmstadt,
  Planckstr. 1, D-64259 Darmstadt, Germany \and
  Institut f\"ur Kernphysik, Technische Universit\"at Darmstadt,
  Schlossgartenstr. 9, D-64289 Darmstadt, Germany \and
  Institut de Recherches Subatomiques, IN2P3-CNRS/Universit\'e Louis
  Pasteur, F-67037 Strasbourg, France}

\date{Received: \today / Revised version: \today}

\abstract{We have performed shell-model calculations of the half-lives
  and neutron-branching probabilities of the r-process waiting point
  nuclei at the magic neutron number $N=82$. These new calculations
  use a larger model space than previous shell model studies and an
  improved residual interaction which is adjusted to recent
  spectroscopic data around $A=130$. Our shell-model results give a
  good account of all experimentally known half-lives and
  $Q_\beta$-values for the $N=82$ r-process waiting point nuclei. Our
  half-life predictions for the $N=82$ nuclei with $Z=42$--46 agree
  well with recent estimates based in the energy-density functional
  method.
}

\maketitle

\section{Introduction}

The astrophysical r-process produces about half of the heavy elements
in the Universe by a sequence of fast neutron-capture reactions
interrupted by photodissociations and followed by $\beta$ decays,
running through extremely neutronrich nuclei far off the valley of
stability \cite{Burbidge.Burbidge.ea:1957,Cameron:1957}.  The $\beta$
decays of the r-process waiting points, associated with nuclei with
magic neutron numbers $N=50, 82$ and 126 play a crucial role for the
r-process dynamics and elemental abundance distributions
\cite{Cowan.Thielemann.Truran:1991}.
Despite their importance only a few half-lives of waiting points with
magic neutron numbers $N=50$ and 82 are known experimentally
\cite{Pfeiffer.Kratz.ea:2001,Hosmer.Schatz.ea:2005}, while no
experimental data exist yet for the $N=126$ waiting points.  This
paper is concerned with the $N=82$ waiting points, and fortunately
here the half-lives of $^{131}$In, $^{130}$Cd and $^{129}$Ag have been
measured and serve as stringent constraints for models which have to
be used to predict the unknown half-lives of the other waiting points
and r-process nuclei.  

Beta-decays are notoriously difficult to model as they are determined
by the weak low-energy tails of the Gamow-Teller strength
distribution, mediated by the operator $\bm{\sigma \tau_-}$.  There
have been several previous estimates for the half-lives of the $N=82$
waiting points based on the Quasiparticle Random Phase Approximation
on top of semi-empirical global models
\cite{Moeller.Nix.Kratz:1997,Borzov.Goriely:2000}, the energy-density
functional (DF3) method \cite{Borzov:2006} or the
Hartree-Fock-Bogoliubov (HFB) method \cite{Engel.Bender.ea:1999}.  A
comparison to the data show that the predicted half-lives are in the
right order of magnitude, but are often somewhat too long.  This might
imply that the models underestimate the correlations among nucleons
which pull down the Gamow-Teller (GT) strength to low excitation
energies. It is well known that the interacting shell model is the
method of choice to describe the Gamow-Teller distribution in nuclei
\cite{Brown.Wildenthal:1988,Caurier.Langanke.ea:1999,Langanke.Martinez-Pinedo:2000}
and in fact, the best agreement with the measured half-lives has been
achieved within a shell model approach
\cite{Martinez-Pinedo.Langanke:1999}.

This shell model calculation as well as the other theoretical
approaches have been challenged recently by the experimental
determination of the excitation energy of the first $1^+$ state in
$^{130}$In, which carries most of the GT$_-$ strength of the
$^{130}$Cd decay \cite{Dillmann.Kratz.ea:2003}.  Here, gamma rays
observed in the beta decay of $^{130}$Cd~\cite{Dillmann.Kratz.ea:2003}
yield an excitation energy of $E_x=2.16$~MeV, in contrast to the shell
model predictions which place this state at noticeably lower energy,
$E_x= 1.4$ MeV \cite{Dillmann.Kratz.ea:2003} and 1.5 MeV
\cite{Martinez-Pinedo.Langanke:1999}.  As the halflife has a strong
energy dependence which approximately scales like $(Q_\beta-E_x)^5$,
where $Q_\beta=M_i-M_f$ is the difference of the masses of the parent
and daughter nuclei, respectively, the misplacement of the $1^+$
excitation energy has been fortuitiously cancelled in the shell model
calculation \cite{Martinez-Pinedo.Langanke:1999} by the use of too
small a $Q_\beta$ value which was taken from the Duflo-Zuker mass
model yielding $7.56$~MeV \cite{Duflo.Zuker:1995}, while the
experimental $Q_\beta$ value is 8.34~MeV.  Thus the shell model
calculation of \cite{Martinez-Pinedo.Langanke:1999} despite its
successful desription of the halflives of the $N=82$ waiting point
nuclei, has to be improved to account also for the recent experimental
structure data concerning the respective decays.  It is the aim of
this manuscript to present such an improved study.

\section{Formalism}

\begin{figure}
  \centering
  \includegraphics[width=\linewidth]{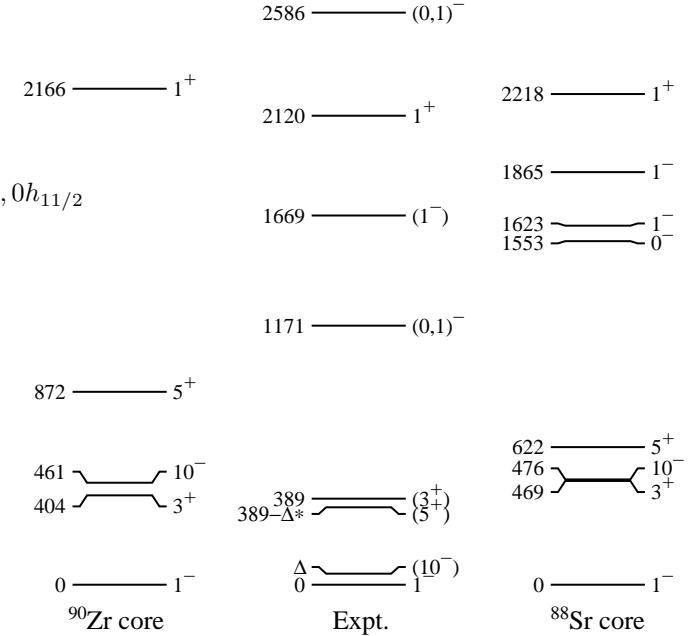} 
  \caption{Comparison of low-energy shell model spectrum for
    $^{130}$In with the data. The shell model spectrum of the left is
    calculated for the model space assuming a $^{90}$Zr core, the one
    on the right for the model space with a $^{88}$Sr core. Details of
    the calculations are given in the text.\label{fig:in130}}
\end{figure}

We have performed large-scale shell model calculations using the code
ANTOINE \cite{Antoine}. We have improved the previous shell model
study of the half-lives of the $N=82$ r-process
nuclei~\cite{Martinez-Pinedo.Langanke:1999} in several ways. At first,
we used a larger model space which includes the
$0g_{7/2},1d_{3/2,5/2},2s_{1/2},0h_{11/2}$ orbitals outside the $N=40$
core for neutrons, thus assuming a closed $N=82$ shell configuration
in the parent nucleus. For protons our model space was spanned by the
$0g_{9/2,7/2},1d_{3/2,5/2},2s_{1/2}$ orbitals.  Thus our model space
avoids spurious center-of-mass excitations by omitting the $h_{11/2}$
orbit for protons and the $g_{9/2}$ orbit for neutrons. Secondly, we
have performed the calculations of the parent ground states and the GT
strength distributions in the daughter nucleus for the $N=82$ parents
with charge numbers $Z=43-49$ including all possible correlations
within the defined model space, a clear improvement with respect to
the calculations in ref.~\cite{Martinez-Pinedo.Langanke:1999}. But
most importantly we have modified the residual interaction adopted in
our studies to reproduce relevant experimental nuclear structure
information.  These modifications were guided by the observation that
for the nuclei of interest here, the halflives are dominated by
Gamow-Teller transitions to states at low excitation energy in the
daughter nuclei and that these transitions are mainly determined by a
single transition matrix element in which a $g_{7/2}$ neutron is
changed into a $g_{9/2}$ proton. This implies that the transition
matrix elements are quite insensitive to the relative position of the
$g_{7/2}$ neutron orbit but its position determines the excitation
energy of the daughter states and the $Q_\beta$ value for the
evaluation of the half-lives. Starting from the interaction used in
the previous shell-model
calculations~\cite{Martinez-Pinedo.Langanke:1999} we have implemented
several monopole modifications aiming to reproduce the known $Q_\beta$
values of $^{131}$In~\cite{Fogelberg.Gausemel.ea:2004} and
$^{130}$Cd~\cite{Dillmann.Kratz.ea:2003}, the position of the
$h_{11/2}$ neutron orbit~\cite{Fogelberg.Gausemel.ea:2004} and the
experimental excitation energy of the first $1^+$ state in $^{130}$In
(for details see ref.~\cite{gniady.others:2007}). This has resulted in
two effective interactions: one that allows proton excitations from
the $p_{1/2}$ orbit (that means uses a $^{88}$Sr core) and another one
where these excitations are supressed (uses a $^{90}$Zr
core). Fig.~\ref{fig:in130} compares our calculated low-energy
spectrum of $^{130}$In with the data and find good agreement for the
excitation energies of the first $3^+$ state. Importantly our improved
shell model calculations also reproduces the unexpectedly high
excitation energy of the first $1^+$ state, which is of key importance
for the calculation of the $^{130}$Cd halflife.  However, the low
energy $5^+$ state and the states with possible assigments $0^-$ and
$1^-$ are missed by the calculation that uses a $^{90}$Zr core. These
are reproduced by the interactions that includes the $p_{1/2}$ orbital
which is energetically relatively close to the $g_{9/2}$ orbital, but
carries the opposite parity. Once, the $p_{1/2}$ orbital is included
in the model space we are forced to truncate the number of protons
excited across the $g_{9/2}$ shell gap. The calculations shown at the
right of figure~\ref{fig:in130} where performed allowing for 6 protons
to be excited from the $p_{1/2}$ and $g_{9/2}$ orbitals. Shell model
calculations performed within the same model space and using the same
interaction also reproduce the recently measured spectrum of
$^{130}$Cd~\cite{Jungclaus.Caceres.ea:2007}. It is particularly
noteworthy that this includes the excitation energy of the first
excited $2^+$ state which is calculated at $E_x=1.325$ MeV in close
agreement with the experimental value of 1.346 MeV, hence not
confirming the tentative suggestion that this state would reside at
$E_x=0.957$ MeV which had been interpreted as an onset of shell
quenching already in $^{130}$Cd~\cite{Kautzsch.Walters.ea:2000}.

As stated above, our calculation of the halflives are based on the
valence space on top of the $^{90}$Zr as this allows for untruncated
calculations and the changes in the computed values of the half-lives
are negligible if we adopt the model space with the $^{88}$Sr core
(and the appropriate interaction).

\section{Results}

\begin{figure}
  \centering
  \includegraphics[width=\linewidth]{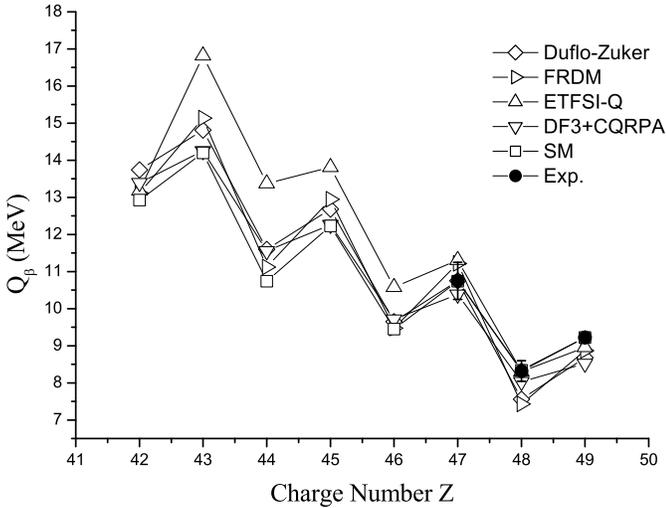} 
  \caption{Comparison of shell model $Q_\beta$ values for the $N=82$
    isotones to the Audi-Wapstra systematics
    \cite{Audi.Wapstra.Thibault:2003} and to predictions of other
    models: FRDM \cite{Moeller.Nix.Kratz:1997}, ETFSI-Q
    \cite{Pearson.Nayak.Goriely:1996}, DF3 \cite{Borzov:2006} 
and Duflo-Zuker \cite{Duflo.Zuker:1995}. 
\label{fig:Qvalue}}
\end{figure}

As discussed above, $\beta$ half-lives depend sensitively on the
$Q_\beta$ value. Unfortunately this important quantity is
experimentally not known for most of the relevant $N=82$ r-process
nuclei and hence has to be estimated based on theoretical models or by
systematic extrapolations from data for neighboring nuclei.  Our
$Q_\beta$ values were calculated from the energies of the isobaric
analog states
with a systematic correction for the Coulomb displacement
energies~\cite{Antony.Pape.Britz:1997}.  In Fig. \ref{fig:Qvalue} we
compare the shell model $Q_\beta$-values with experiment, the
Audi-Wapstra systematics~\cite{Audi.Wapstra.Thibault:2003} and other
theoretical calculations. As stated above, our shell model calculation
is tuned by proper monopole modifications to reproduce the
experimental $Q_\beta$ value for $^{130}$Cd. However, we find also a
good agreement to the other $Q_\beta$ values as given in the
compilation of Audi-Wapstra \cite{Audi.Wapstra.Thibault:2003}. Thus as
the shell model $Q_\beta$ values are close to the experimental values
and agree with those of most other models within the theoretical
uncertainties, we will in the following adopt the shell model
$Q_\beta$ values in our calculation of the half-lives.  However, the
exception are the predictions of the quenched Extended Thomas-Fermi
with Strutinski Integral (ETFSI-Q)
model~\cite{Pearson.Nayak.Goriely:1996} which predicts
$Q_\beta$-values which are noticeably larger than those of the other
models, most noticeably for the most neutron deficient $N=82$ isotones.

\begin{figure}
  \centering
  \includegraphics[width=\linewidth]{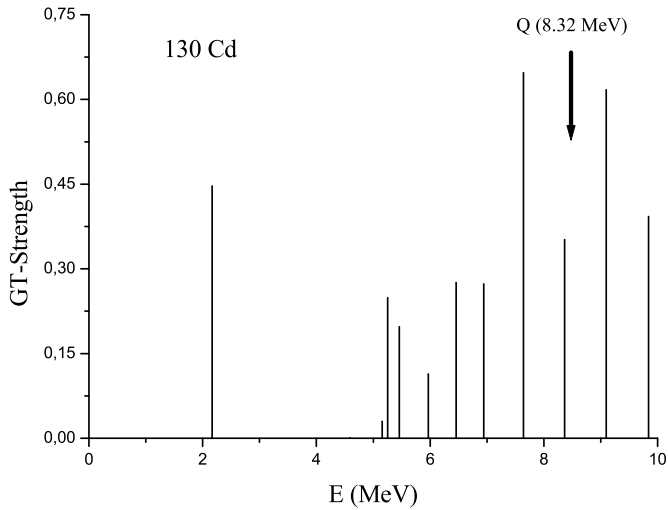} 
  \caption{GT$_-$ strength distribution for $^{130}$Cd.
\label{fig:Gtstrength_Cd}}
\end{figure}

\begin{figure}
  \centering
  \includegraphics[width=\linewidth]{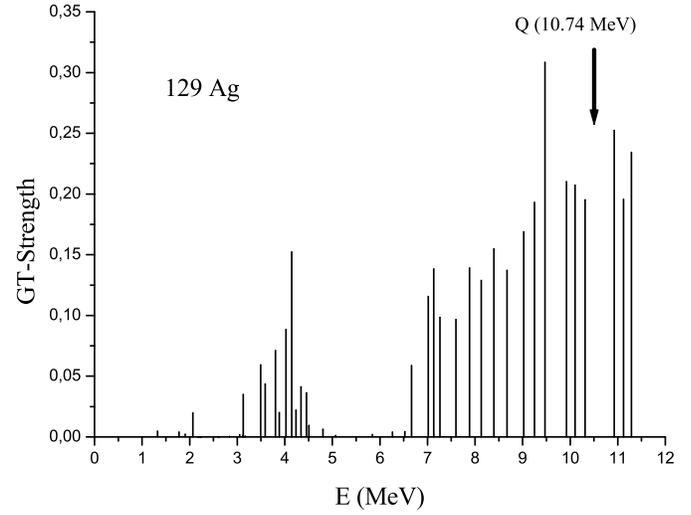} 
  \caption{GT$_-$ strength distribution for $^{129}$Ag.
\label{fig:Gtstrength_Ag}}
\end{figure}

\begin{figure}
  \centering
  \includegraphics[width=\linewidth]{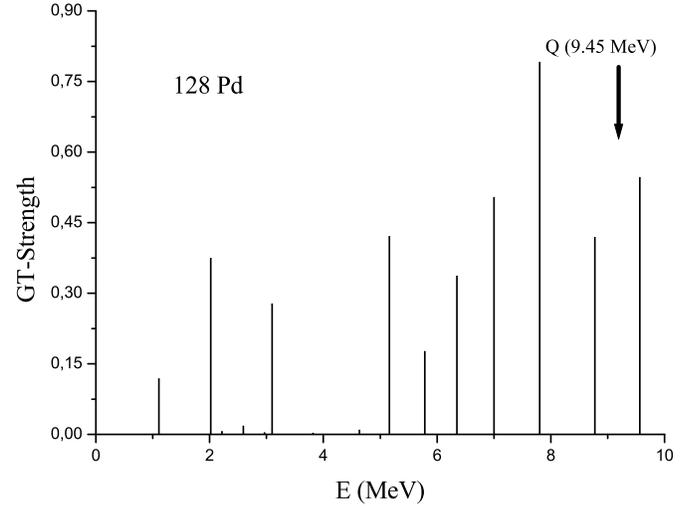} 
  \caption{GT$_-$ strength distribution for $^{128}$Pd.
\label{fig:Gtstrength_Pd}}
\end{figure}

\begin{figure}
  \centering
  \includegraphics[width=\linewidth]{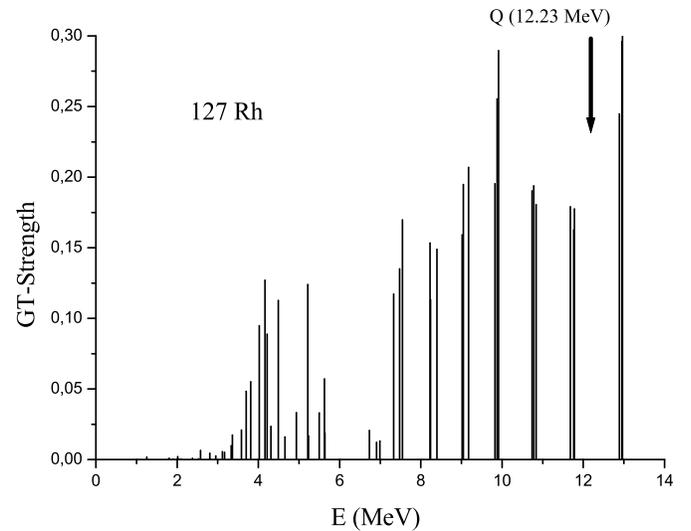} 
  \caption{GT$_-$ strength distribution for $^{127}$Rh.
\label{fig:Gtstrength_Rh}}
\end{figure}

\begin{figure}
  \centering
  \includegraphics[width=\linewidth]{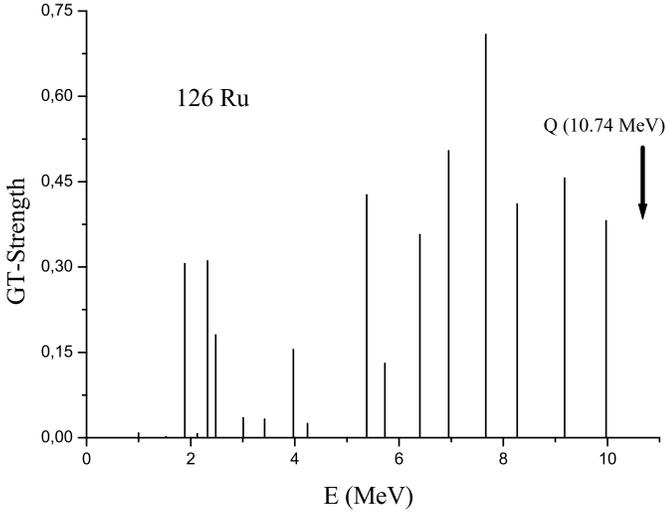} 
  \caption{GT$_-$ strength distribution for $^{126}$Ru.
\label{fig:Gtstrength_Ru}}
\end{figure}

\begin{figure}
  \centering
  \includegraphics[width=\linewidth]{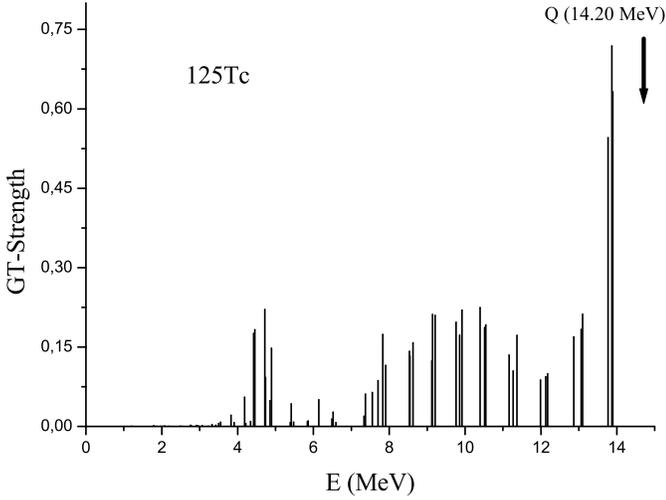} 
  \caption{GT$_-$ strength distribution for $^{125}$Tc.
\label{fig:Gtstrength_Tc}}
\end{figure}

\begin{figure}
  \centering
  \includegraphics[width=\linewidth]{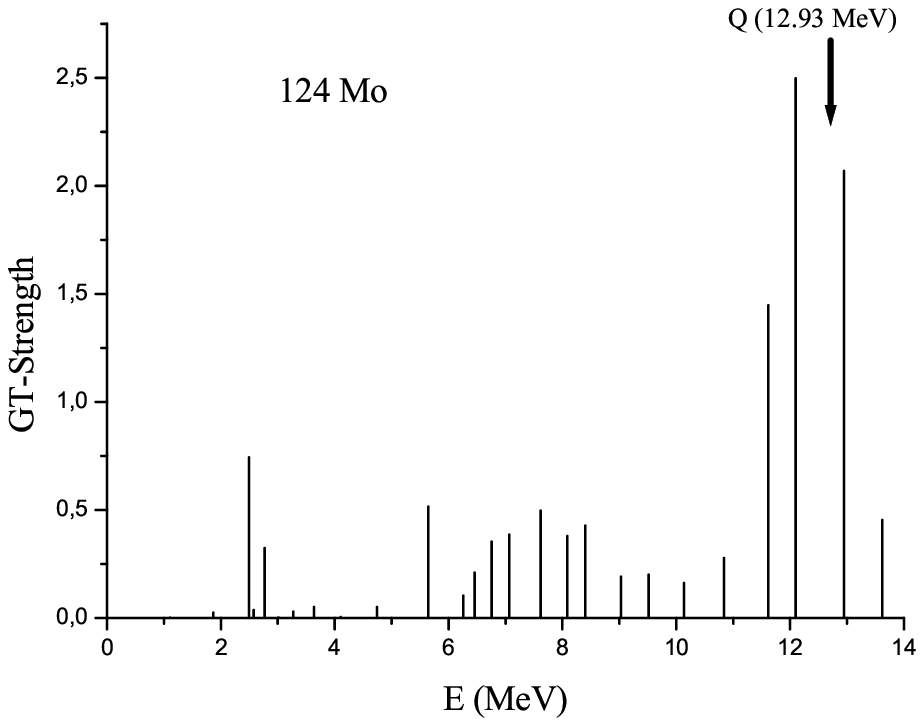} 
  \caption{GT$_-$ strength distribution for $^{124}$Mo.
\label{fig:Gtstrength_Mo}}
\end{figure}

The final ingredients of our half-life calculations are the GT$_-$ strength
functions. We have calculated them within our shell model approach
using the Lanczos method with 60 iterations for each possible final
$J$-value. The calculated  GT$_-$ strength functions for the
$N=82$ nuclei from $^{124}$Mo ($Z=42$) to $^{131}$In ($Z=49$) are shown
in Figs.
\ref{fig:Gtstrength_Cd}-\ref{fig:Gtstrength_Mo}.  For even-even
nuclei the GT transitions lead to $J_f^\pi=1^+$ states in the
daughters, while for the odd-$A$ nuclei the final states can have $J_f
= J_i-1, J_i, J_i+1$, where $J_i$ is the angular momentum of the
parent ground state. For the calculation of the half-lives, the choice
of the appropriate Gamow-Teller quenching factor is an important
issue. It is wellknown that shell model calculations reproduce the GT
strength distributions (total strength and fragmentation) very well
within complete 0$\hbar\omega$ calculations (i.e. model spaces which
include a complete major oscillator shell), if the GT operator is
quenched by a constant factor.  This factor has been determined for
$sd$ shell~\cite{Brown.Wildenthal:1985,Brown.Wildenthal:1988}, where
it is 0.77, and the $pf$ shell, where it is
0.74~\cite{Martinez-Pinedo.Poves.ea:1996b}. Unfortunately the
appropriate constant for nuclei in the mass $A=130$ region has not yet
been determined. Thus we adopt a quenching factor of 0.71 adjusted to
reproduce the experimental half-life of $^{130}$Cd ($t_{1/2}= 162\pm
7$ ms).  

Here one word about first-forbidden transitions is in order. While
a study by M\"oller and collaborators, based on the FRDM/QRPA for 
GT transitions and the statistical gross theory for forbidden theories, 
indicates sizable contributions of first-forbidden transitions to the
half-lives, the only consistent microscopic treatment of GT and first-forbidden
transitions for the $N=82$ half-lives by Borzov, based on the energy
density-functional method, implies that forbidden transitions
accelerate the halflives only slightly by about $10\%$ or less.
Based on the later result our halflives calculated purely from GT transitions
are meaningful, as  a small, but roughly constant contribution
of first-forbidden transition can be absorbed into a modification of the
quenching factor which would be $0.73$ (very close to the standard
quenching factor), if first-forbidden transitions
yield a $10\%$ contribution.

Our shell model 
half-life for $^{131}$In (260 ms) agrees also with the measured values
($280\pm30$ s). However, we overestimate the one for $^{129}$Ag slightly
(70 ms to be compared with $46^{+5}_{-9}$
s)~\cite{Kratz.Others:1998,Pfeiffer.Kratz.ea:2001}. The present
shell-model half-lives are longer for nuclei with $Z\leq 47$ than the
ones computed in reference~\cite{Martinez-Pinedo.Langanke:1999} (see
table~\ref{tab:lives}). There are two reasons for the change in
half-lives. First, for nuclei with $Z=42$--44 the $Q_\beta$-values
obtained in the present shell-model calculations are around 1~MeV
smaller than the Duflo-Zuker values used in
ref.~\cite{Martinez-Pinedo.Langanke:1999} (see fig.~\ref{fig:Qvalue}).
Second, for all the nuclei the low lying Gamow-Teller strength is
shifted upwards around 0.5~MeV in excitation energy.
This shift results in
longer half-lives even if the $Q_\beta$-value were unmodified as it is
the case in $^{129}$Ag. This shift is due to the monopole
modifications introduced in the interaction used for the present
calculations to increase the excitation energy of the $1^+$ state in
$^{130}$In from 1.6~MeV~\cite{Martinez-Pinedo.Langanke:1999} to the
experimental value of 2.12~MeV~\cite{Dillmann.Kratz.ea:2003}.

\begin{figure}
  \centering
  \includegraphics[width=\linewidth]{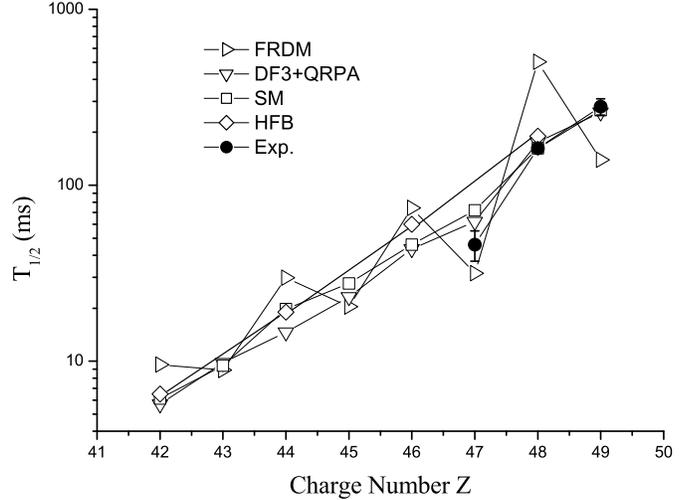} 
  \caption{Comparison of half-lives of the $N=82$ isotones as
    calculated in the FRDM \cite{Moeller.Nix.Kratz:1997}, HFB
    \cite{Engel.Bender.ea:1999}, and the present shell model
    approaches with data
    \cite{Pfeiffer.Kratz.ea:2001,Dillmann.Kratz.ea:2003}.\label{fig:halflives}}
\end{figure}

\begin{table}
  \caption{Comparison of the present shell model half-lives and the
    ones of reference~\cite{Martinez-Pinedo.Langanke:1999} with
    experiment~\cite{Kratz.Others:1998,Pfeiffer.Kratz.ea:2001,%
      Dillmann.Kratz.ea:2003}. All   
    half-lives are in ms.}  
  \label{tab:lives}
  \renewcommand{\arraystretch}{1.1}
    \begin{tabular}{cccc} \hline\hline
    Nucleus & \multicolumn{3}{c}{Half-Life (ms)} \\ \cline{2-4}
    &   Expt.  & \textrm{Theor.} & \textrm{Theor.
    (ref~\cite{Martinez-Pinedo.Langanke:1999})} \\ \hline 
    $^{131}$In  & $280 \pm 30$ & 260   &   177  \\
    $^{130}$Cd  & $162 \pm 7$  & 162   &    146  \\
    $^{129}$Ag  & $46^{+5}_{-9}$ & 70   &    35.1     \\
    $^{128}$Pd  &              & 46    &   27.3  \\
    $^{127}$Rh  &              & 27.65 &    11.8  \\
    $^{126}$Ru  &              & 19.76 &    9.6  \\
    $^{125}$Tc  &              & 9.44  &   4.3  \\
    $^{124}$Mo  &              & 6.13  &     3.5   \\
    \hline\hline
  \end{tabular}
\end{table}

\begin{figure}
  \centering
  \includegraphics[width=\linewidth]{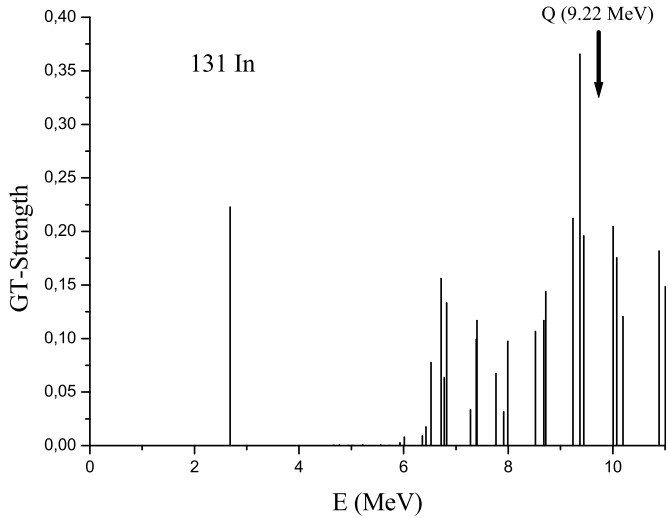} 
  \caption{GT$_-$ strength distribution for $^{131}$In.
\label{fig:Gtstrength_In}}
\end{figure}

In Fig. \ref{fig:halflives} and Table~\ref{tab:lives} our results are
compared with those of other theoretical models and the data.  The
current half-lives are quite similar to those obtained within the HFB
model of \cite{Engel.Bender.ea:1999} and are also in good agreement
with the DF3-QRPA approach~\cite{Borzov:2003}. The latter is reassuring as
the DF3-QRPA model also reproduces the half-lives of other spherical
nuclei in the vicinity of $N=82$ quite well~\cite{Borzov.Others:2007}.
Our half-lives show a mild odd-even effect, with the half-lives of the
even-even waiting point nuclei slighly enlarged with respect to the
neighboring odd-A nuclei.  This is caused by a partial cancellation of
the odd-even staggering in the $Q_\beta$ values (Fig.
\ref{fig:Qvalue}) by the larger excitation energies of the lowest GT
states in the odd-$A$ daughter nuclei (Figs.
\ref{fig:Gtstrength_Cd}-\ref{fig:Gtstrength_Mo}).

As shown in Fig. \ref{fig:Qvalue} the various models
predict $Q_\beta$ values within a variation of about 1 MeV, which, due
to the strong energy dependence of the phase space, translates into
the largest uncertainties of the $\beta$ half-lives. To estimate this
uncertainty we have recalculated the half-lives by replacing the shell
model $Q_\beta$ values by those of the different models and find
half-lives which agree with the present ones within a factor of two or
better, except for the larger ETFSI-Q $Q_\beta$ values which result in
significantly smaller half-lives for the most proton-deficient nuclei.
It is also worth noting that by replacing our shell model $Q_\beta$ value
for $^{129}$Ag by the systematic Audi-Wapstra value, we find agreement with the experimental
half-life within the uncertainties of the systematic $Q_\beta$ value.

\begin{figure}
  \centering
  \includegraphics[width=\linewidth]{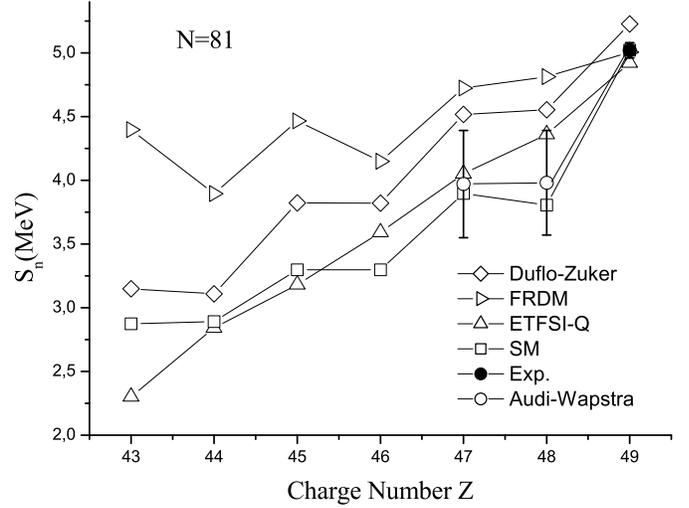} 
  \caption{Comparison of the shell model neutron separation energies
    for the $N=81$ isotones to the data
    \cite{Audi.Wapstra.Thibault:2003} and predictions of other models:
    FRDM \cite{Moeller.Nix.Kratz:1997}, Duflo-Zuker
    \cite{Duflo.Zuker:1995} and
    ETFSI-Q~\cite{Pearson.Nayak.Goriely:1996}.
\label{fig:Sn}} 
\end{figure}

\begin{figure}
  \centering
  \includegraphics[width=\linewidth]{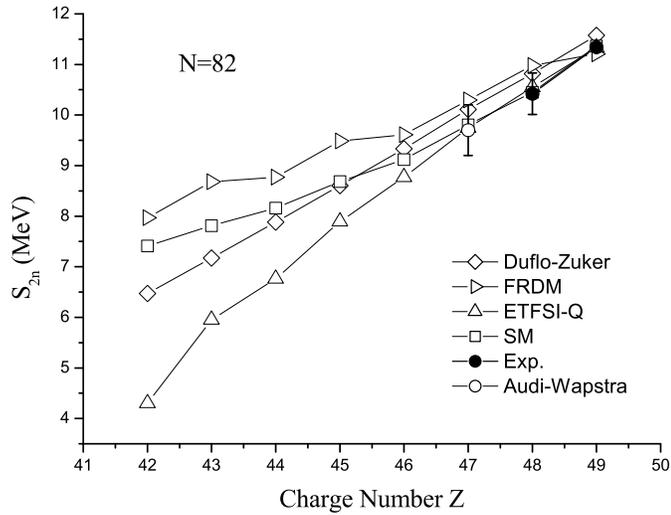} 
  \caption{Comparison of the shell model two-neutron separation
    energies for the $N=82$ 
    nuclei to the data \cite{Audi.Wapstra.Thibault:2003} and predictions of other
    models: FRDM \cite{Moeller.Nix.Kratz:1997}, Duflo-Zuker \cite{Duflo.Zuker:1995}
and ETFSI-Q \cite{Pearson.Nayak.Goriely:1996}.
\label{fig:S2n}}
\end{figure}

As the neutron separation energies in the daughter nuclei are quite small,
some of the GT$_-$ strengths resides actually at energies above
the neutron emission threshold and $\beta$ decays to these states
are followed by neutron emission. The probability for $\beta$-delayed
neutron emission depends sensitively on the neutron separation
energies $S_n$ which are not known experimentally for most of the
nuclei of interest here and have to be estimated theoretically. 
In Fig. \ref{fig:Sn} we compare our shell model neutron separation energies
for the $N=81$ daughter nuclei with those of various models. 
Importantly for
$^{131}$Sn and $^{130}$In the neutron separation energies are known 
experimentally and our present results agree quite nicely with the data.
The neutron separation energies predicted in the FRDM model 
\cite{Moeller.Nix.Kratz:1997} are larger
than all other model predictions and also exceed the experimental
values for 
$^{131}$Sn and $^{130}$In.  
Our shell model results predict a quite similar
slope of the $S_n$ values as found 
in the Duflo-Zuker mass model \cite{Duflo.Zuker:1995}
yielding quite sizable
neutron separation energies even in the proton-deficient nuclei $^{125}$Ru
and $^{124}$Tc. This is in difference to the ETFSI-Q
values \cite{Pearson.Nayak.Goriely:1996} which predict a pronounced weakening of the $S_n$ values
with increasing neutron excess. A quite similar behavior is found
if one compares the 2-neutron separation energies of the $N=82$
r-process nuclei (Fig. \ref{fig:S2n}). Again, our shell model
results agree with the available data ($^{131}$In and $^{130}$Cd)
and show a significantly slower decrease of the $S_{2n}$ values
than predicted by the ETFSI model, while the FRDM and Duflo-Zuker
models yield 2-neutron separation energies which agree reasonably well
with the shell model ones.

\begin{figure}
  \centering
  \includegraphics[width=\linewidth]{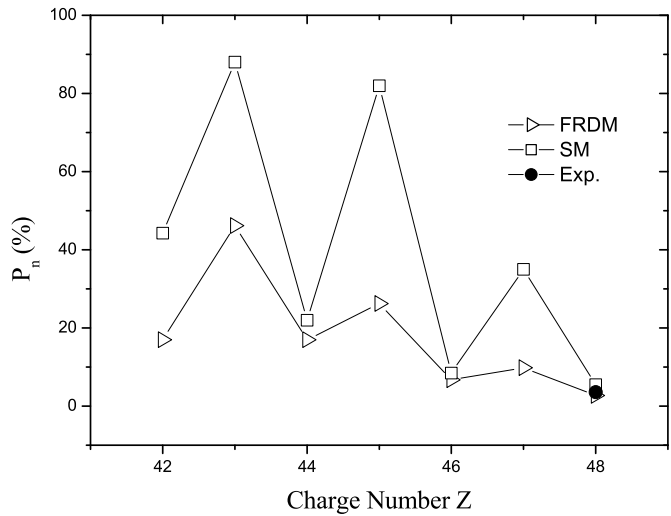} 
  \caption{One neutron emission probabilities.\label{fig:P1n}}
\end{figure}

Using our
shell model GT$_-$ strength functions and $S_n$ values we have calculated the
probability $P_n$ that the $\beta$ decay is accompanied by the
emission of (at least) one neutron, defined as the relative probability of the
$\beta$-decay rate above the neutron emission threshold $S_n$.  
The results are shown in Fig. \ref{fig:P1n}, indicating that within
our shell model study most of the $\beta$ decays go to states
below the neutron threshold for the even-even parent nuclei, 
yielding probabilities for
$\beta$-delayed neutron emission of 40\% or less.
The shell model predicts quite a strong odd-even staggering in
the $P_n$ values, indicating $P_n$ values of $80\%$ or larger for
$^{127}$Rh and $^{125}$Tc. The FRDM model \cite{Moeller.Nix.Kratz:1997}
also predicts an odd-even dependence in the neutron emission
probabilities, however, this effect is somewhat smaller than
for the shell model values. For $^{130}$Cd, the $P_n$ value is known
experimentally \cite{Dillmann.Kratz.ea:2003} and it agrees with the FRDM and
shell model predictions.

\section{Conclusion}

We have recalculated shell model half-lives and neutron
emission probabilities for the $N=82$ waiting point nuclei in the
r-process, improving a previous shell model study by enlargement
of the model space and by modification of the residual interaction
which reproduces recent spectroscopic findings for nuclei in this
regime of the nuclear chart. In particular our modified calculation
reproduces the unexpectedly high excitation energy of the first
$1^+$ state in $^{130}$In \cite{Dillmann.Kratz.ea:2003}. This, as a good
description of the $Q_\beta$ value, is crucial as the GT transition 
to this low-lying state dominates the $^{130}$Cd half-life. 
We 
find good agreement with the experimentally known
half-lives for the $Z=48,49$ nuclei and overestimate the
one of $^{129}$Ag slightly. 

As in most other models
\cite{Moeller.Nix.Kratz:1997,Engel.Bender.ea:1999,Borzov.Goriely:2000}
the shell model half-lives have been computed solely on the assumption
of an allowed Gamow-Teller transition. The importance of forbidden
transitions for the half-lives of the $N=82$ waiting point nuclei is
somewhat controversial. Using the gross theory for the forbidden
transitions, M\"oller \cite{Moeller.Pfeiffer.Kratz:2003} finds
non-negligible contributions, while Borzov, adopting the QRPA approach
within the Fermi-liquid theory, concludes that forbidden transitions
play only a minor role \cite{Borzov:2003}. In view of these
differences, shell model calculations of the forbidden $\beta$ decays
of the $N=82$ r-process waiting point nuclei are desirable.

\begin{acknowledgement}
  I. N. Borzov is supported by a grant from the German DFG under
  contract number 436 RUS 113/907/0-1.
\end{acknowledgement}


\end{document}